\documentclass[a4paper]{jpconf}
\usepackage{graphicx}

\begin{document}

\title{IceCube PeV Neutrinos and Leptophilic Dark Matter}

\author{Marco Chianese}

\address{Dipartimento di Fisica {\it Ettore Pancini}, Universit\`a di Napoli Federico II, Complesso Univ. Monte S. Angelo, I-80126 Napoli, Italy}
\address{INFN, Sezione di Napoli, Complesso Univ. Monte S. Angelo, I-80126 Napoli, Italy}

\ead{chianese@na.infn.it}

\begin{abstract}
We analyze the scenario where the IceCube high energy neutrino events are explained in terms of an extraterrestrial flux due to two different components: a contribution coming from know astrophysical sources for energies up to few hundreds TeV and a top-down contribution originated by the decay of heavy dark matter particles with a mass of few PeV. Contrary to previous approaches, we consider a leptophilic three-body decay that dominates at PeV energies due to the absence of quarks in the final state. We find that the theoretical predictions of such a scenario are in a slightly better agreement with the IceCube data if the astrophysical component has a cut-off at about 100~TeV. This interpretation of IceCube data can be easily tested in the near future since the decaying dark matter scenario predicts a sharp cut-off at PeV energy scale and the observation of an anisotropy towards Galactic Center of our Galaxy in contrast with the isotropic astrophysical flux.
\end{abstract}

\section{Introduction}

The nature of Dark Matter (DM) still represents a mystery that physicists try to unveil since more than 80 years from its first evidence in the Coma galaxy cluster by Fritz Zwicky. A lot of interesting schemes have been provided in order to allocate DM candidates with very different masses, from about $10^{-32}$ up to $10^{18}$~GeV. The most promising scenario is the Weakly Interacting Massive Particle (WIMP) paradigm, in which the interaction between the Standard Model (SM) and the DM is of the order of the weak interactions and the DM particle can have a mass from $\mathcal{O}(1)$~GeV up to $\mathcal{O}(100)$~TeV due to the unitary constraint \cite{Griest:1989wd}. However, the missing evidence of a DM particle in all direct searches (DM production at colliders as LHC and SM--DM scattering) suggests that the only viable way to obtain information about very massive DM candidates would exploit indirect searches in astrophysical observations.

The IceCube (IC) Neutrino Observatory \cite{Aartsen:2013jdh,Aartsen:2014gkd} represents a good experiment to observe high energy phenomena where massive DM particles produce neutrinos with very high energy. After three years of data-taking (2010-2013) IC experiment has collected 37 neutrino events with deposited energy from 30~TeV up to 2~PeV. Since the compatibility with the expected atmospheric background (muons and neutrinos produced by the decay of $\pi$ and $K$ and prompt neutrinos coming from the decay of charmed mesons) is excluded at 5.7~$\sigma$, the origin of such events has to be related to some extraterrestrial processes.

In the paper \cite{Boucenna:2015tra} we have analyzed the possibility that the PeV neutrinos are produced by leptophilic three-body decays of DM particles. Contrary to previous scenarios \cite{DM}, we assume that the IC neutrino spectrum can be explained by the sum of three different components:
\begin{itemize}
\item the {\it atmospheric} component, which dominates for energies up to 60~TeV;
\item a bottom-up component (hereafter denoted as {\it astrophysical neutrino flux}) from know astrophysical sources in the 60 - 300~TeV energy range;
\item a top-down additional component (hereafter denoted as {\it DM neutrino flux}) for higher energy.
\end{itemize}
Therefore, the total neutrino flux for $E_\nu \geq 60$~TeV is equal to
\begin{equation}
\frac{{\rm d} J}{{\rm d}E_\nu} \left(E_\nu\right) = \frac{{\rm d} J_{\rm Ast}}{{\rm d}E_\nu} \left(E_\nu\right) +  \frac{{\rm d} J_{\rm DM}}{{\rm d}E_\nu} \left(E_\nu\right) \,.
\label{eq:flux_sum}
\end{equation}
The number of neutrino events for a given energy bin $\left[ E_i,\,E_{i+1} \right]$ can be obtained by the relation
\begin{equation}
N_i = 4 \pi \Delta t \int^{E_{i+1}}_{E_i} {\rm d}E \sum_{\alpha=e,\mu,\tau} \frac{{\rm d}J^\alpha_{\nu+\overline{\nu}}}{{\rm d}E} \,A_\alpha \left(E\right) \,,
\label{eq:number_neutrino}
\end{equation}
where $\Delta t = 988$~days is the exposure time and $A_\alpha \left(E\right)$ is the neutrino effective area for different neutrino flavour $\alpha$ \cite{Aartsen:2013jdh}. Due to the low statistics at our disposal, we have considered for simplicity the equivalence between deposited and neutrino energy. Such an approximation does not affect dramatically our qualitative results \cite{Boucenna:2015tra}.

\section{Astrophysical neutrino flux}

There exist many astrophysical sources that are potentially able to produce high energy neutrinos through the acceleration of protons, which then interact with themselves ($pp$ interactions) or with photons ($p\gamma $ interactions). The most popular astrophysical sources proposed for IceCube are the extragalactic Supernova Remnants (SNR) \cite{Chakraborty:2015sta}, the Active Galactic Nuclei (AGN) \cite{Kalashev:2014vya,Stecker:1991vm} and the Gamma Ray Bursts (GRB) \cite{Waxman:1997ti}. All these bottom-up scenarios are affected by large uncertainties since there is not a complete knowledge about the physics of the proton acceleration mechanism. Moreover, each astrophysical source is not able to fit alone all the IC observations. For instance, extragalactic SNR has a cut-off in energy of the order of 100~TeV, whereas AGN provides a good description of IC neutrino flux at high energy only.

In the present analysis, in order to cover all the acceleration mechanisms related to different astrophysical sources, we parametrize the astrophysical component of the total neutrino flux either with a Unbroken Power Law (UPL) 
\begin{equation}
E_\nu^2 \frac{{\rm d} J_{\rm Ast}}{{\rm d}E_\nu} \left(E_\nu\right) = J_0 \,\left(\frac{E_\nu}{100\,  {\rm TeV}}\right)^{-\gamma} \,,
\label{eq_UPL_def}
\end{equation}
or with a Broken Power Law (BPL), characterized by an exponential cut-off at some energy scale $E_0$
\begin{equation}
E_\nu^2 \frac{{\rm d} J_{\rm Ast}}{{\rm d}E_\nu} \left(E_\nu\right)  = J_0 \,\left(\frac{E_\nu}{100\,  {\rm TeV}}\right)^{-\gamma}\, \exp{\left( -\frac{E_\nu}{E_0}\right)} \,.
\label{eq_BPL_def}
\end{equation}
In the previous expressions, the quantity $\gamma+2$ is the spectral index while $J_0$ is the normalization of the flux. As suggested by the extragalactic SNR results \cite{Chakraborty:2015sta}, we fix the value of the cut-off $E_0$ to be equal to 125~TeV.

\section{Dark Matter neutrino flux}

In our analysis we study the case of a heavy fermionic singlet DM candidate $\chi$, which decays into SM particles through a coupling having few characteristics. In particular, we require that
\begin{itemize}
\item the DM particle is directly coupled to neutrinos via a leptophilic coupling. Indeed any coupling to other SM particles (Higgs, gauge bosons and quarks) would unavoidably lead to an abundant production of secondary neutrinos, ruling out any astrophysical component;
\item the coupling is non-renormalizable in order to suppress the lifetime of $\chi$ with powers of the energy scale of new physics. This improves the need of an unnatural tiny coupling that in general has to be $\mathcal{O}\left(10^{-30}\right)$;
\item the DM particle decays into a multi-body final state. In this way, the DM neutrino flux is not peaked at a given energy but is spread to lower energy.
\end{itemize}
The first operator that satisfies all the previous requirements is the {\it non-renormalizable lepton portal}
\begin{equation}
\frac{y_{\alpha\beta\gamma}}{M_{\rm Pl}^2} \left(\overline{L_\alpha}  {\ell}_\beta\right) \left( \overline{L_\gamma} \chi \right)+ \mathrm{h.c.} \, ,
\label{eq:DMop}
\end{equation}
where $L$ is the lepton left-handed doublet, $\ell$ is the right-handed lepton and $\{\alpha,\beta,\gamma\}$ are flavour indices. We assume that the energy scale of new physics is the Planck mass $M_{\rm Pl}$. In order to forbid the other operators as $\overline{L}\tilde{H}\chi$, we invoke a global flavour symmetry as $U_f(1)$, for which with a suitable choice of the charges we have $\{\alpha,\beta,\gamma\}\equiv \{\mu, e,\tau \}+\{\tau,e, \mu \}+\{e, \mu, e \}$ (see Ref.~\cite{Boucenna:2015tra} for more details about this model and the other case of non-Abelian groups). Assuming for simplicity that $y \equiv \left| y_{\mu e \tau} - y_{\tau e \mu} \right|=\left| y_{e \mu e} \right|$, the lifetime of the DM particle $\chi$, having a mass $M_\chi$, is equal to 
\begin{equation}
\tau_\chi^{-1} = \frac{3 y^2}{6144 \, \pi^3} \frac{M_\chi^5}{M_{\rm Pl}^4} \, .
\end{equation}

In this framework, the top-down component to the neutrino flux consists in a galactic contribution (G) and an extragalactic one (EG), whose expressions are given by 
\begin{eqnarray}
\frac{{\rm d} J_{\chi}^{\rm G}}{{\rm d}E_\nu} \left(E_\nu,l,b\right) & = & 
\frac{1}{4\pi\,M_{\chi}\,\tau_{\chi}} 
\sum_{\alpha=e,\mu,\tau} \frac{{\rm d}N^\alpha_{\nu + \bar{\nu}}}{{\rm d}E_\nu}\left(E_\nu\right)
\int_0^\infty {\rm d}s\; 
\rho_{\chi}(r(s,l,b)) \, , \\
\frac{{\rm d}J^{\rm EG}_{\chi}}{{\rm d}E_\nu} \left(E_\nu,l,b\right) & = &
\frac{\Omega_{\chi}\rho_{\rm cr}}{4\pi M_{\chi} \tau_{\chi}}
\int_0^\infty {\rm d}z\,
\frac{1}{H(z)}
\sum_{\alpha=e,\mu,\tau} \frac{{\rm d}N^\alpha_{\nu + \bar{\nu}}}{{\rm d}E_\nu}\left((1+z)E_\nu\right) \,,
\end{eqnarray}
where $\rho_{\chi}(r)$ is the Navarro-Frenk-White density profile\footnote{The analysis is mostly independent on the choice of the density profile (Einasto, Isothermal, etc.).} of DM particles in our Galaxy as a function of distance $r$ from the Galactic center, $H(z)=H_0 \sqrt{\Omega_\Lambda+\Omega_{\rm m}(1+z)^3}$ is the Hubble expansion rate as a function of redshift $z$ and $\rho_{\rm cr}=5.5\times10^{-6}\,{\rm GeV}\, {\rm cm}^{-3}$ is the critical density of the Universe. In the analysis, we consider the $\Lambda$CDM cosmology where $\Omega_\Lambda=0.6825$, $\Omega_{\rm m}=0.3175$, $\Omega_{\chi}=0.2685$ and $h\equiv H_0/100\,{\rm km}\,{\rm s}^{-1}\,{\rm Mpc}^{-1}=0.6711$ (from the Planck experiment \cite{Ade:2015xua}). Finally, the quantity ${\rm d}N^\alpha_{\nu + \bar{\nu}}/{\rm d}E_\nu$ is the neutrino energy spectrum coming from the decay of $\chi$. It has been evaluated by means of a MonteCarlo procedure, taking into account the electroweak radiative corrections that in general change the energy spectrum at energies almost two orders of magnitude smaller than the DM mass \cite{Ciafaloni:2011sa}. It is worth observing that our results satisfy the Fermi-LAT bound on the total electromagnetic energy \cite{Ackermann:2014usa}.

\section{Results}

The analysis has been done by using a multi-Poisson likelihood fit. The chi-square takes the form
\begin{equation}
\chi^2=-2 \ln {\mathcal{L}}=2 \sum_i \left[N_i - n_i + n_i \ln\left(\frac{n_i}{N_i}\right) \right] \,,
\end{equation}
where $N_i$ is the expected number of neutrinos for energy bin (Eq.~\ref{eq:number_neutrino}), while $n_i$ is the observed one. The quantity $\gamma$ has been varied in the physical range [0, 1], whereas the DM mass $M_\chi$ has been restricted to the range [1~PeV,~10~PeV], finding the best-fit values to be equal to $\gamma=1.0$ for UPL case, $\gamma=0.0$ for BPL one, and $M_\chi=5.0$~PeV in both models.
\begin{figure}[t!]
\begin{center}
\includegraphics[width=0.46\textwidth]{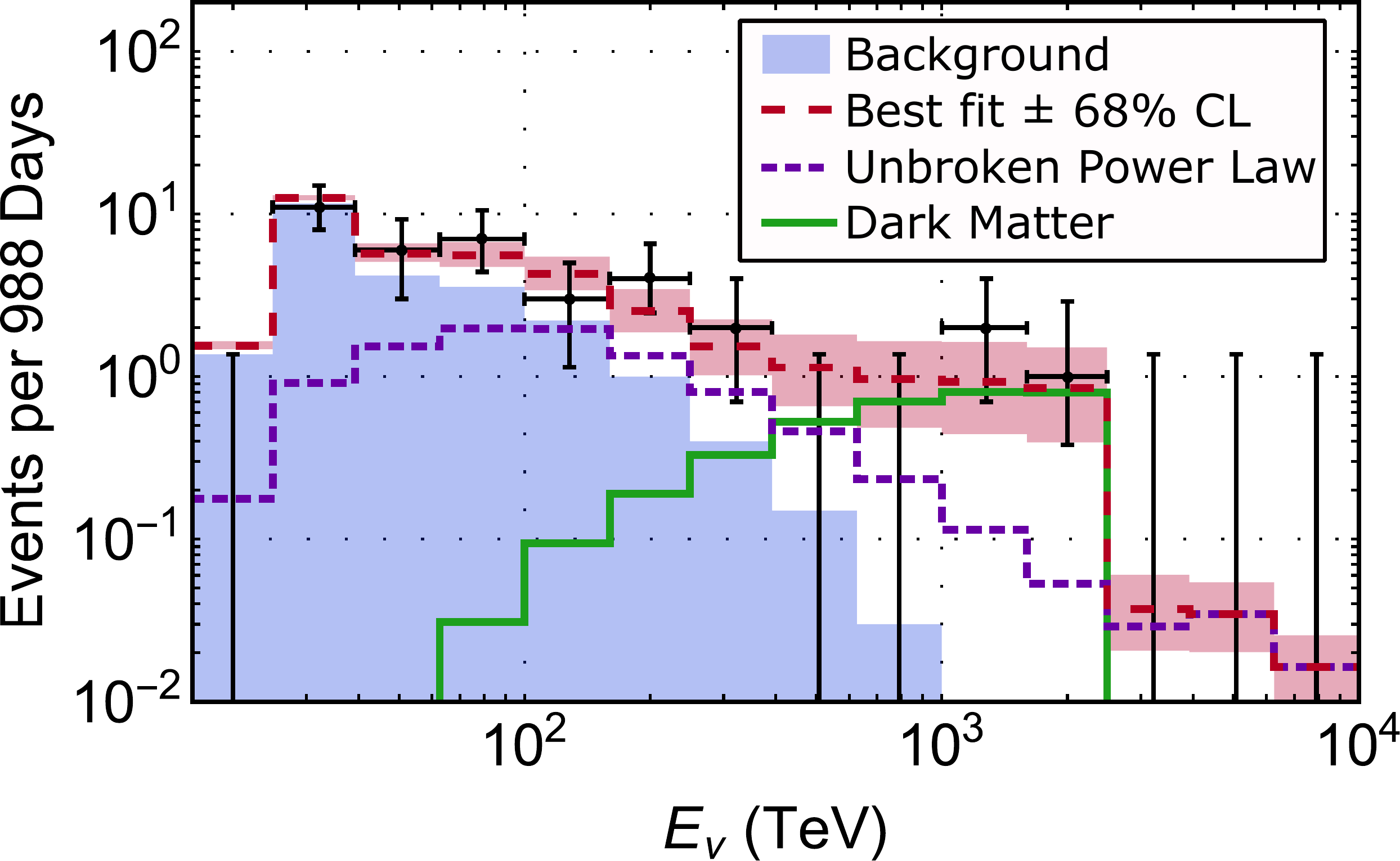}
\hskip5.mm
\includegraphics[width=0.46\textwidth]{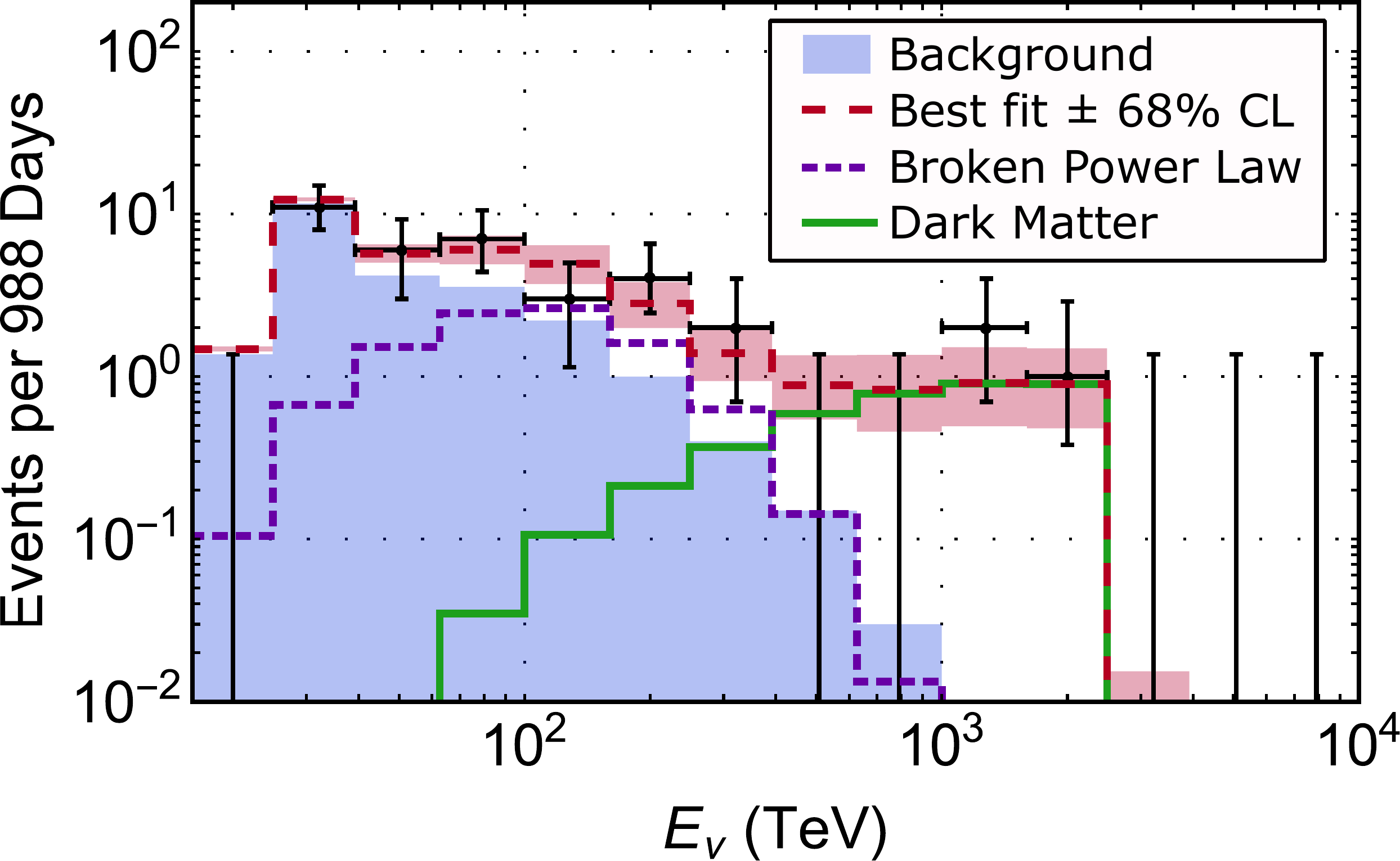}
\includegraphics[width=0.4\textwidth]{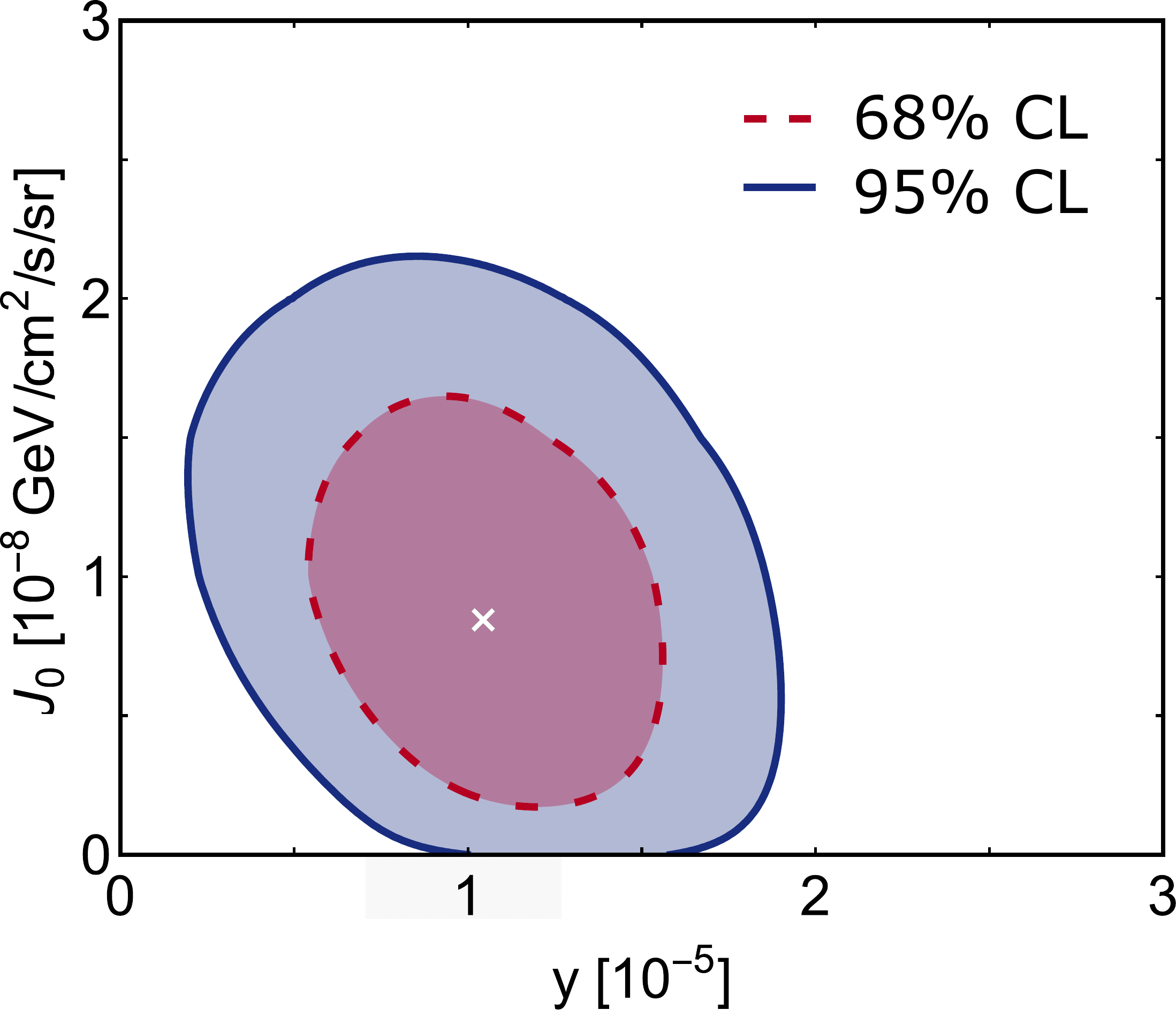}
\hskip5.mm
\includegraphics[width=0.4\textwidth]{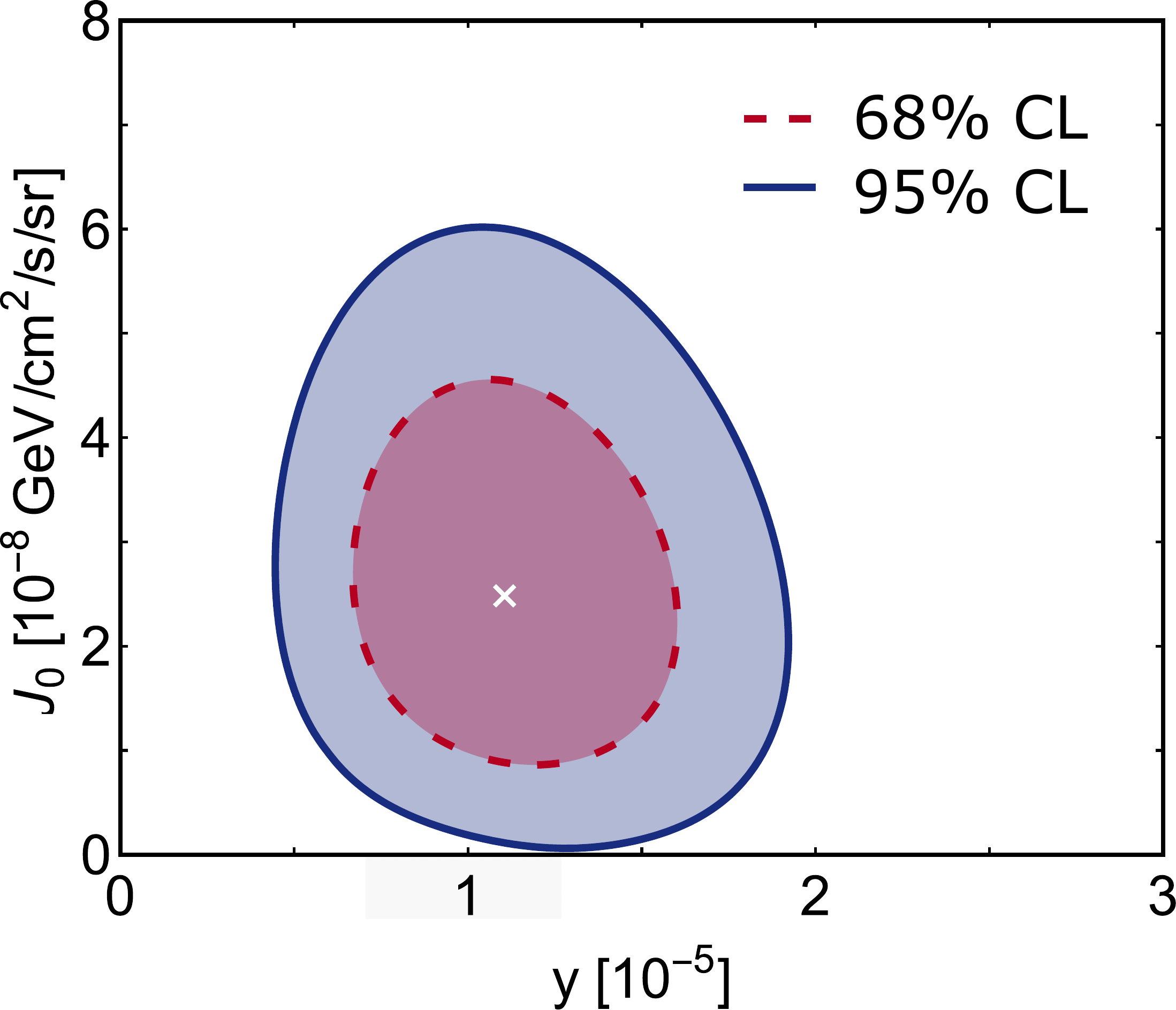}
\\(A)\hskip69.mm (B)\\
\caption{In the first row we report the neutrino events as a function of the neutrino energy $E_\nu$ for the DM+UPL (column A)  and  DM+BPL (column B) models. The red (long-dashed) line is the best fit, i.e. the sum of atmospheric (blue region), astrophysical (purple-dashed) and DM (green-solid) components, and the black points represent the IC data. The second row shows the 68\% C.L. (dashed) and 95\% C.L. (solid) contours for the two parameters $y$ and $J_0$ corresponding to DM+UPL (column A) and DM+BPL (column B). The crosses are the best-fit points.}
\label{fig:U1}
\end{center}
\end{figure}
In the upper part of Fig.~\ref{fig:U1}, we report the fit of neutrino events for both models (DM+UPL and DM+BPL). In the lower part, we show the 68\% C.L. and 95\% C.L. contours for the SM--DM coupling $y$ and the normalization of astrophysical flux $J_0$. In particular, the marginalized 95\% C.L. best-fit values are
\begin{eqnarray}
{\rm UPL}: & \quad & y \, [10^{-5}] = 1.0^{+0.7}_{-0.7}\,, \qquad J_0\, [10^{-8}] = 0.8^{+1.0}_{-0.5}\,; \\
&&\nonumber\\
{\rm BPL}: & \quad & y \, [10^{-5}] = 1.1^{+0.6}_{-0.5}\,, \qquad J_0\, [10^{-8}] = 2.5^{+2.8}_{-2.1}\,;
\end{eqnarray}
where $J_0$ is expressed in unit of GeV cm$^{-2}$ s$^{-1}$ sr$^{-1}$. It is worth observing that in both schemes the DM component provides a non-vanishing contribution at $2\sigma$ level. As we can see from the values of reduced chi-square, the IC data slightly prefer the BPL parametrization ($\chi^2/{\rm dof}=9.2/12$) rather than the UPL one ($\chi^2/{\rm dof}=10.3/12$). In Ref.~\cite{Boucenna:2015tra} we report also the results for the case of non-Abelian flavour symmetry $A_4$, even if the two schemes do not show a significant difference.

\section{Conclusions}

The IceCube Neutrino Observatory had the first evidence of extraterrestrial high energy neutrinos, which can be related to the presence of new physics. We have analyzed the scenario where the IC neutrino flux corresponds to the sum of three different contributions: the atmospheric background ($E_\nu \leq 60$~TeV), an astrophysical component (60~TeV$\leq E_\nu\leq300$~TeV) and a top-down component ($E_\nu \geq 300$~TeV) that arises from the three-body decay of a leptophilic DM candidate with a mass of 5.0~PeV. Even though the IC neutrino flux can be explained only by an astrophysical origin due to the low statistics, the decaying DM scenario is very intriguing since it can provide important information on DM physics and can give indication on the direction for future DM experiments. Moreover, it is worth observing that such a scenario can be easily tested in the early future. Indeed, it predicts the presence of a sharp cut-off above few PeV and an anisotropy towards the Galactic Center because the DM neutrino flux consists in a galactic contribution (almost 2/3 of the total flux) and an extragalactic one.

Even if our results regard only the IC three years data, the decaying DM scenario is still in agreement with the new IC data of four years \cite{Aartsen:2015zva}. A new track event with a deposited energy of about 2.6~PeV has been observed, but the energy of the primary neutrino is unknown because such an event is not fully contained in the IC detector. However, if the neutrino energy was of the order of few PeV, this would not change our scenario once the DM mass is slightly shifted to high energy.

\section*{\bf Acknowledgments}

We acknowledge support by the Instituto Nazionale di Fisica Nucleare I.S. TASP and the PRIN 2012 �Theoretical Astroparticle Physics� of the Italian Ministero dell�Istruzione, Universit\`a e Ricerca.

\end{document}